\documentstyle[12pt]{article}
\catcode`\@=11

\@addtoreset{equation}{section}
\def\theequation{\arabic{section}.\arabic{equation}}

\catcode`\@=11

\def\section{\@startsection{section}{1}{\z@}{3.5ex plus 1ex minus
   .2ex}{2.3ex plus .2ex}{\large\bf}}

\def\eqnarray{\stepcounter{equation}\let\@currentlabel=\theequation
    \global\@eqnswtrue
    \global\@eqcnt\z@\tabskip\@centering\let\\=\@eqncr
    $$\halign to \displaywidth\bgroup\@eqnsel\hskip\@centering
      $\displaystyle\tabskip\z@{##}$&\global\@eqcnt\@ne
       \hfil${{}##{}}$\hfil
      &\global\@eqcnt\tw@ $\displaystyle\tabskip\z@{##}$\hfil
       \tabskip\@centering&\llap{##}\tabskip\z@\cr}
\def\lefteqn#1{\hbox to 4\arraycolsep{$\displaystyle #1$\hss}}
\def\thesection{\arabic{section}.}

\def\appendix{\setcounter{section}{0}
        \def\thesection{Appendix}
        \def\theequation{\Alph{section}.\arabic{equation}}}

\long\def\@makefntext#1{\parindent 0cm\noindent
\hbox to 1em{\hss$^{\@thefnmark}$}#1}
\def\IR{{\hbox{{\rm I}\kern-.2em\hbox{\rm R}}}}
\def\IH{{\hbox{{\rm I}\kern-.2em\hbox{\rm H}}}}
\def\IC{{\ \hbox{{\rm I}\kern-.6em\hbox{\bf C}}}}
\def\IZ{{\hbox{{\rm Z}\kern-.4em\hbox{\rm Z}}}}
\def\rref#1{(\ref{#1})}
\newcommand{\beq}{\begin{equation}}
\newcommand{\eeq}{\end{equation}}
\begin{document}
%
%
%
%
\def\citen#1{%
\edef\@tempa{\@ignspaftercomma,#1, \@end, }
\edef\@tempa{\expandafter\@ignendcommas\@tempa\@end}%
\if@filesw \immediate \write \@auxout {\string \citation {\@tempa}}\fi
\@tempcntb\m@ne \let\@h@ld\relax \let\@citea\@empty
\@for \@citeb:=\@tempa\do {\@cmpresscites}%
\@h@ld}
%
\def\@ignspaftercomma#1, {\ifx\@end#1\@empty\else
   #1,\expandafter\@ignspaftercomma\fi}
\def\@ignendcommas,#1,\@end{#1}
%
%
\def\@cmpresscites{%
 \expandafter\let \expandafter\@B@citeB \csname b@\@citeb \endcsname
 \ifx\@B@citeB\relax 
    \@h@ld\@citea\@tempcntb\m@ne{\bf ?}%
    \@warning {Citation `\@citeb ' on page \thepage \space undefined}%
 \else
    \@tempcnta\@tempcntb \advance\@tempcnta\@ne
    \setbox\z@\hbox\bgroup 
    \ifnum\z@<0\@B@citeB \relax
       \egroup \@tempcntb\@B@citeB \relax
       \else \egroup \@tempcntb\m@ne \fi
    \ifnum\@tempcnta=\@tempcntb 
       \ifx\@h@ld\relax 
          \edef \@h@ld{\@citea\@B@citeB}%
       \else 
          \edef\@h@ld{\hbox{--}\penalty\@highpenalty \@B@citeB}%
       \fi
    \else   
       \@h@ld \@citea \@B@citeB \let\@h@ld\relax
 \fi\fi%
 \let\@citea\@citepunct
}
%
\def\@citepunct{,\penalty\@highpenalty}%
%
%
\def\@citex[#1]#2{\@cite{\citen{#2}}{#1}}%
%
%
\def\@cite#1#2{\leavevmode\unskip
  \ifnum\lastpenalty=\z@ \penalty\@highpenalty \fi 
  \ [{\multiply\@highpenalty 3 #1
      \if@tempswa,\penalty\@highpenalty\ #2\fi 
    }]\spacefactor\@m}
\let\nocitecount\relax  
%
\begin{titlepage}
\vspace{.5in}
\begin{flushright}
UCD-92-30\\
December 1992\\
gr-qc/9301006
\end{flushright}
\vspace{.5in}
\begin{center}
{\Large\bf
Real Tunneling Solutions\\ and the Hartle-Hawking Wave Function}\\
\vspace{.4in}
{S.~C{\sc arlip}\footnote{\it email: carlip@dirac.ucdavis.edu}\\
       {\small\it Department of Physics}\\
       {\small\it University of California}\\
       {\small\it Davis, CA 95616}\\{\small\it USA}}
\end{center}

\vspace{.5in}
\begin{center}
{\large\bf Abstract}
\end{center}
{\small
A real tunneling solution is an instanton for the Hartle-Hawking
path integral with vanishing extrinsic curvature (vanishing ``momentum'')
at the boundary.  Since the final momentum is fixed, its conjugate
cannot be specified freely; consequently, such an instanton will contribute
to the wave function at only one or a few isolated spatial geometries.  I
show that these geometries are the extrema of the Hartle-Hawking wave
function in the semiclassical approximation, and provide some evidence
that with a suitable choice of time parameter, these extrema are the maxima
of the wave function at a fixed time.
}
\end{titlepage}
\addtocounter{footnote}{-1}

\section{Introduction}

In the Hartle-Hawking approach to quantum cosmology \cite{Hawking,HartHawk},
the wave function of the universe is obtained from a Euclidean path
integral over metrics and matter fields on a manifold $M$ with a single
boundary component $\partial M$.  Such a path integral depends on the
induced metric $h_{ab}$ and the matter configuration $\varphi|_{\partial M}$
on ${\partial M}$, thus determining a functional
\beq
\Psi[h,\varphi|_{\partial M}] =
  \sum_M \int [dg][d\varphi]\, \exp\left\{-I_E[g,\varphi]\right\} ,
\label{1x1}
\eeq
where the summation represents a sum over topologies.  The functional
$\Psi$ is the Hartle-Hawking wave function, which we are instructed to
interpret as an amplitude for finding a universe characterized by $h$
and $\varphi|_{\partial M}$.  The Hartle-Hawking approach neatly finesses
the question of initial conditions by omitting any initial boundary,
and it postpones the question of the nature of time in quantum gravity:
information about time is hidden in the boundary geometry $h$, but the
path integral can be formulated without making a choice of time explicit.

In references \cite{entropy} and \cite{sum}, it was shown that the sum
over topologies in \rref{1x1} can have a drastic effect on the behavior
of the wave function.  For most of this paper, this complication will
be ignored; we shall focus on the contribution of a single fixed topology
$M$ to the wave function of the universe.  In particular, we shall look
for the maxima of $\Psi$, the most probable spatial geometries and matter
configurations of the universe.

In the saddle point approximation, the wave function \rref{1x1} takes the
standard form
\beq
\Psi_M[h,\varphi|_{\partial M}] \sim \sum_{\{\bar g,\bar\varphi\}}
  D_M[\bar g, \bar\varphi]\exp\left\{-I_E[\bar g,\bar\varphi]\right\} ,
\label{1x2}
\eeq
where the the sum is now over classical solutions $\{\bar g,\bar\varphi\}$
with the prescribed boundary values, and the prefactor $D_M$ is a
combination of determinants coming from gauge-fixing and from small
fluctuations around the extrema.  In this approximation, it may not
always be possible to specify $h$ and $\varphi|_{\partial M}$ arbitrarily
--- for a given topology $M$, some boundary values may not lead to
classical solutions --- but we certainly have a great deal of flexibility.
For instance, given one classical solution, we can cut off a small
neighborhood of the boundary to obtain a new solution with the same
topology but different boundary data.  A given manifold $M$ will
therefore contribute to the wave function over a wide range of values
of $h$ and $\varphi|_{\partial M}$.

A particularly interesting class of saddle points consists of the
so-called real tunneling geometries \cite{GibHart}, solutions of the
field equations for which the metric has Riemannian signature in $M$
and vanishing extrinsic curvature at the boundary $\partial M$.
Classically, these are the solutions for which the metric can be
smoothly joined to a metric with {\em Lorentzian\/} signature to
the future of ${\partial M}$.\footnote{``Smoothly'' here means ``with
finite action.''}  Real tunneling geometries clearly give the right
classical picture of the Hartle-Hawking ``no boundary'' boundary
condition, describing universes characterized by a Lorentzian metric
``now'' but no initial boundary. Moreover, by analogy with more
tractable problems such as pair production in a constant electric
field \cite{Vilenkin}, we expect these solutions to play an important
role in the overall behavior of the path integral.

Note, however, that a real tunneling solution will typically contribute
to $\Psi[h,\varphi|_{\partial M}]$ at only a few isolated values of $h_{ab}$.
In the Hamiltonian formulation of general relativity, the extrinsic
curvature is canonically conjugate to $h_{ab}$, and we should not expect
to be able to freely specify $h_{ab}$ while simultaneously maintaining
$K_{ab}=0$.  This conclusion can be checked explicitly in three dimensions
with $\varphi=0$ and $\Lambda<0$: a real tunneling solution is then a
hyperbolic three-manifold with a totally geodesic boundary, and for any
given topology $M$, the boundary metric $h_{ab}$ either does not exist
(if there is no classical extremum) or is unique.

We shall see below that the boundary values of real tunneling solutions are
in fact the extrema of the wave function $\Psi$.  In general, these extrema
are saddle points rather than maxima.  But we shall find some evidence that
with a suitable choice of time parameter, the real tunneling solutions
determine the genuine maxima of the Hartle-Hawking wave function at a
fixed time.

\section{Scalar Fields}

Before tackling the more difficult problem of gravity, it is instructive
to look at the behavior of the ``Hartle-Hawking wave function'' for a free
scalar field.  The Euclidean action for a field $\varphi$ in $n$ spacetime
dimensions is
\beq
I_E[\varphi] = {1\over2}\int_M d^n\!x\,\sqrt{g}
  \left(g^{ab}\partial_a\varphi\partial_b\varphi + m^2\varphi^2\right) ,
\label{2x1}
\eeq
where for the moment the metric $g$ is fixed.  The first variation of the
action is
\beq
\delta I_E[\varphi]
 = -\int_M d^n\!x\,\sqrt{g}\,(\delta\varphi)(\Delta_g-m^2)\varphi +
 \int_{\partial M}d^{n-1}\!x\,\sqrt{h}\, (\delta\varphi)\nabla_{\!n}\varphi,
\label{2x2}
\eeq
where $\nabla_{\!n}$ is the normal derivative at $\partial M$ and $\Delta_g
= \nabla_{\!a}\nabla^a$ is the Laplacian.  Solutions of the classical
equations of motion are thus genuine extrema as long as $\varphi$ is fixed
at the boundary.

For a free scalar field, the path integral \rref{1x1} is trivial;
the saddle point approximation \rref{1x2} is exact, with $D_M =
\det(\Delta_g-m^2)^{-1/2}$.  In this context, equation \rref{2x2} now
has a new interpretation: it tells us that if we vary the boundary data
$\varphi|_{\partial M}$, the classical action $I_E[\bar\varphi]$ is extremal
when $\nabla_{\!n}\bar\varphi$ vanishes.  This is the scalar analog of
the condition for a real tunneling solution --- a classical solution
$\bar\varphi$ has a finite-action extension across ${\partial M}$ to a
Lorentzian spacetime precisely when $\nabla_{\!n}\bar\varphi = 0$.  We
thus see from \rref{1x2} that the scalar Hartle-Hawking wave function
has its extrema at precisely the boundary values of the ``scalar
real tunneling solutions.''

We can now ask whether these extrema are maxima.  If we restrict our
attention to the space of classical solutions, the second variation of
the Euclidean action, evaluated at the point $\nabla_{\!n}\bar\varphi
= 0$, is
\beq
\delta^2I_E[\varphi] = \int_{\partial M}d^{n-1}\!x\,\sqrt{h}\,
   (\delta\varphi)\nabla_{\!n}(\delta\varphi)
   = \int_M d^n\!x\,\sqrt{g}
   \left(g^{ab}\partial_a(\delta\varphi)\partial_b(\delta\varphi)
   + (\delta\varphi)\Delta_g(\delta\varphi)\right) .
\label{2x4}
\eeq
But for variations in the space of classical solutions,
\beq
(\Delta_g - m^2)(\delta\varphi) = 0 ,
\label{2x5}
\eeq
so
\beq
\delta^2I_E[\varphi]
 = \int_M d^n\!x\,
 \sqrt{g}\left(g^{ab}\partial_a(\delta\varphi)\partial_b(\delta\varphi)
 + m^2(\delta\varphi)^2\right) > 0 .
\label{2x6}
\eeq
Scalar real tunneling solutions are therefore minima of the Euclidean
action, and hence maxima of the wave function.

The scalar Hartle-Hawking wave function is thus peaked at the boundary
value $\varphi|_{\partial M}$ for which the extension $\bar\varphi$ to $M$
has a vanishing normal derivative.  The map from $\varphi|_{\partial M}$ to
$\nabla_{\!n}\bar\varphi$ is known as the Poisson map \cite{Forman,CCDD},
and we can summarize our results by stating that the maxima of the
Hartle-Hawking wave function occur at the zeros of the Poisson map.  For
the more complicated case of gravity, the analog to $\varphi|_{\partial M}$
is the boundary metric $h_{ab}$, while the normal derivative $\nabla_{\!n}
\bar\varphi$ corresponds at least roughly to the extrinsic curvature
$K_{ab}$, and the Poisson map takes the metric boundary data to the
corresponding extrinsic curvature.  Let us now ask whether our results
for the scalar field can be extended to this situation.

\section{Gravity}

For simplicity, we shall concentrate on the case of pure gravity, with a
nonvanishing cosmological constant acting as a stand-in for matter fields.
The Euclidean action $I_E$ is
\beq
I_E[g] = -{1\over16\pi G}\int_M d^n\!x\, \sqrt{g}\,(R[g]-2\Lambda)
         -{1\over8\pi G}\int_{\partial M} d^{n-1}\!x\, \sqrt{h} K ,
\label{3x1}
\eeq
where $R[g]$ is the scalar curvature, $\Lambda$ is the cosmological
constant, and $K$ is the trace of the intrinsic curvature of $\partial M$.
Let $n^a$ be a unit vector field in a neighborhood of $\partial M$ whose
restriction to $\partial M$ is the outward unit normal; for simplicity,
we can choose $n^a$ to satisfy
\beq
n^b\nabla_{\!b}n^a = 0
\label{3x2}
\eeq
near $\partial M$.  By a slight abuse of notation, let
\beq
h_{ab} = g_{ab} - n_an_b .
\label{3x3}
\eeq
(The restriction of $h_{ab}$ to $\partial M$ is the induced spatial metric
on the boundary.)  An easy extension of Wald's calculation in \cite{Wald}
then gives
\begin{eqnarray}
-16\pi G \delta I_E[g] &=&
   \int_M d^n\!x\,\sqrt{g}\,(G_{ab}+\Lambda g_{ab})\delta g^{ab}\nonumber\\
   &+& \int_{\partial M} d^{n-1}\!x\,\sqrt{h}\, \left[
   (K_{ab} - h_{ab}K)\delta g^{ab} + h^{ab}\nabla_{\!a}(n^c\delta g_{bc}
   + 2 g_{bc}\delta n^c)\right] .
\label{3x4}
\end{eqnarray}

Now consider the vector $v_b = n^c\delta g_{bc} + 2 g_{bc}\delta n^c$ in
a neighborhood of $\partial M$.  Using the fact that $n^a$ is a unit
vector, one may easily check that $n^av_a=0$, so $v_a$ is tangential on
$\partial M$.  Thus
\beq
h^{ab}\nabla_{\!a}v_b = D_av^a ,
\label{3x5}
\eeq
where $D_a$ denotes the covariant derivative on $\partial M$ \cite{Wald}.
The last term in \rref{3x4} is therefore the integral of a total derivative
on $\partial M$, and hence vanishes, giving
\beq
-16\pi G \delta I_E[g] =
   \int_M d^n\!x\,\sqrt{g}\,(G_{ab}+\Lambda g_{ab})\delta g^{ab}
   + \int_{\partial M} d^{n-1}\!x\,\sqrt{h}\, (K_{ab} - h_{ab}K)\delta g^{ab}
   ,
\label{3x6}
\eeq
the gravitational version of equation \rref{2x2}.

As in the scalar case, this relation tells us that solutions of the
classical equations of motion are genuine extrema as long as the metric
is fixed on $\partial M$.  But as in the scalar case, it also tells us
more: if we vary the boundary data, the extrema of the classical action
are precisely the real tunneling solutions, for which $K_{ab}=0$ on
$\partial M$.  For the scalar field, the saddle point approximation was
exact, and these extrema were also the extrema of the Hartle-Hawking
wave function.  For gravity, this is no longer true --- the prefactor
$D_M$ in \rref{1x2} can depend nontrivially on the boundary data, and
there will be higher loop corrections --- but as a first approximation,
the real tunneling solutions will again be the extrema of $\Psi[h]$, with
corrections suppressed by powers of Planck's constant.

To determine whether these extrema are maxima, we must again compute the
second variation of the Euclidean action at the point $K_{ab}=0$:
\beq
16\pi G \delta^2I_E[g] = \int_{\partial M} d^{n-1}\!x\,\sqrt{h}\,
   \gamma^{ab}\delta(K_{ab} - h_{ab}K) ,
\label{3x7}
\eeq
where the variation of the metric has been denoted by $\delta g_{ab} =
\gamma_{ab}$ (note that $\delta g^{ab} = -\gamma^{ab}$).  A set of rather
routine calculations is described in the Appendix; the end result is that
with the gauge choice $(n^a\gamma_{ab}h^b{}_c)|_{\partial M} = 0$,
\beq
32\pi G \delta^2I_E[g] = \int_{\partial M} d^{n-1}\!x\,\sqrt{h}\,\left[
 \gamma^{ab}\nabla_{\!n}\gamma_{ab} - {1\over2}\,\gamma\nabla_{\!n}\gamma
 + n^a\beta_a(\gamma - 2\gamma_{ab}n^an^b)\right] ,
\label{3x8}
\eeq
where $\gamma = g^{ab}\gamma_{ab}$ and $\beta^a = \nabla_{\!b}(\gamma^{ab}
- {1\over2}g^{ab}\gamma)$.  This relation is of the same general form as
\rref{2x4}, but in contrast to the scalar case, it is not manifestly
positive.  In particular, let us assume that we can consistently choose
harmonic gauge, $\beta_a = 0$ (see the Appendix for a brief discussion of
this choice).  Then using Lichnerowicz's results for the variation of the
Ricci tensor \cite{Lich},
\beq
\delta R_{ab} = -{1\over2} \Delta_g\gamma_{ab} + {1\over2}\left(
   \gamma_a{}^cR_{bc} + \gamma_b{}^cR_{ac}\right) - \gamma^{cd}R_{acbd}
   = {2\Lambda\over n-2}\gamma_{ab} ,
\label{3x9}
\eeq
we obtain
\begin{eqnarray}
32\pi G \delta^2I_E[g] &=& \int_M d^n\!x\,\sqrt{g}\,\biggl[
  \nabla_{\!c}\gamma_{ab}\nabla^c\gamma^{ab}
  - {1\over2}\,\nabla_{\!c}\gamma\nabla^c\gamma \nonumber\\
  &-& 2\gamma^{ab}\gamma^{cd}C_{acbd}
  + {4\Lambda\over(n-1)(n-2)}\left(\gamma_{ab}\gamma^{ab} +
  {\textstyle{n-3\over2}}\gamma^2\right)
  \biggr] ,
\label{3x10}
\end{eqnarray}
which has no clear positivity properties.

Indeed, it is easy to see that the boundary value of a real tunneling
solution is generically a saddle point of the Hartle-Hawking wave
function, and not a maximum.  For an empty space solution of the
gravitational field equations with $\Lambda\ne0$, the Euclidean action
$I_E$ is proportional to the volume of $M$.  But we can always decrease
this volume by cutting off a small neighborhood of the boundary
$\partial M$, or increase it by extending $M$ slightly past $\partial M$.

Intuitively, though, such a movement of the boundary is not the kind of
deformation we should be concerned about.  It is well known that the
spatial geometry described by the metric $h_{ab}$ implicitly contains
information about time \cite{Wheeler}, and a movement of the entire
boundary $\partial M$ corresponds physically to a time translation.
It is not so surprising that the wave function has no maxima if such
translations are allowed;  the more interesting question is whether
$\Psi[h]$ has any maxima {\em at a fixed time}.

To answer this question, we must extract information about time from
the spatial metric $h_{ab}$ and the extrinsic curvature $K_{ab}$ at
the boundary $\partial M$.  This is not easy to do, since there is
no unique way to parametrize time, but the structure of the real
tunneling geometries suggests one natural approach: we can use York's
``extrinsic time'' $K$ as a time parameter \cite{York}.

If we restrict ourselves to ``fixed time'' variations of the metric, those
for which $K$ remains zero, we can eliminate at least some of the negative
terms from \rref{3x10}.  From equations \rref{a9} and \rref{a12} of the
Appendix, we see that in harmonic gauge, $\delta K=0$ implies that
\beq
\nabla_{\!n} \gamma=0 .
\label{3x10a}
\eeq
On the other hand, equation \rref{3x9} tells us that
\beq
\Delta_g\gamma + {4\Lambda\over (n-2)}\gamma = 0 .
\label{3x11}
\eeq
For a metric $g_{ab}$ with Riemannian signature, the Laplacian $\Delta_g$
with boundary conditions \rref{3x10a} has no positive eigenvalues.
Hence for $\Lambda<0$, $\gamma$ must vanish, eliminating the term
$\nabla_{\!c}\gamma\nabla^c\gamma$ from \rref{3x10}.  For $\Lambda>0$,
equation \rref{3x11} may sometimes have nontrivial solutions, but the
spectrum of the Laplacian is discrete, so there will be at most finitely
many of them; for most variations of the metric, $\gamma$ will again
vanish.

The term in \rref{3x10} involving the Weyl tensor is harder to control.
To temporarily avoid this problem, it is interesting to look at the
simple model of quantum gravity in three spacetime dimensions, where
$C_{abcd}$ is identically zero.  Here, at least, it may be possible to
prove that $\delta^2I_E$ is strictly positive when $\delta K = 0$.

Suppose first that $\Lambda>0$.  We must then check two conditions:
that harmonic gauge $\beta_a=0$, assumed in the derivation of \rref{3x10},
can always be reached, and that \rref{3x11} has no solutions with $\gamma
\ne0$.  If these conditions are satisfied, the second variation \rref{3x10}
will be manifestly positive.

Both conditions require knowledge of the eigenvalues of Laplacians on
$M$, acting on functions in \rref{3x11} and on one-forms in \rref{a6}.
Since these Laplacians are defined in terms of a background metric $g_{ab}$
for which $\partial M$ is totally geodesic, they extend smoothly to the
(closed) double $\widehat M$ of $M$ formed by attaching two copies of $M$
along their boundaries.  The boundary conditions \rref{3x10a} and \rref{a5}
then guarantee that any solution of \rref{3x11} or \rref{a6} extends to
a solution of the corresponding equation on $\widehat M$.  Moreover,
any closed three-dimensional Einstein space with $\Lambda>0$ is a quotient
of the three-sphere by some group of isometries \cite{Wolf}, so it is
sufficient to look at Laplacians on the sphere.  The appropriate
eigenvalues are then given by Ray \cite{Ray}, and it is not hard to
check that both of our conditions are met.

If $\Lambda<0$, our task is somewhat harder, since the last term in
\rref{3x10} is no longer positive.  In this case, the question can be
reformulated in terms of hyperbolic geometry: the Euclidean action
is proportional to the volume of $M$, and the problem is to show that
among hyperbolic metrics on $M$ with $K=0$ on $\partial M$, the metric
for which $\partial M$ is totally geodesic has the smallest volume.
One possible approach is the following.  Hyperbolic structures on $M$
--- metrics with constant curvature $-1$ --- are parametrized by
a finite-dimensional moduli space $\cal M$.  It is plausible that
the volume of $M$ becomes large outside a compact region of $\cal M$,
where the metric on $\partial M$ starts to degenerate.  If this is true,
and if the volume is a reasonably well-behaved function on $\cal M$,
then it must have a minimum.  But among hyperbolic metrics on a fixed
manifold $M$, the real tunneling solution is unique, and is the only
extremum of the volume; it must therefore be the minimum.  It should
be stressed that this argument is at best a strategy for a proof; work
on this question is in progress.

In four dimensions the situation is more complex, since the
Weyl tensor in \rref{3x10} need not vanish.  The inclusion of matter
is likely to provide an added complication.  For the moment, it seems
difficult to make any definitive statements about the maxima of the
Hartle-Hawking wave function for realistic cosmologies.  But the partial
results from three dimensions suggest a conjecture: that the maxima of
the Hartle-Hawking wave function at ``time'' $K=0$ are precisely the
boundary values of the real tunneling solutions.

\section{Conclusion}

The Hartle-Hawking program offers an intriguing approach to quantum
cosmology, but it has not been easy to extract predictions from the
formalism.  Up to now, most work has involved either simple minisuperspace
models (for example, \cite{xx}) or classical or semiclassical
approximations \cite{entropy,sum,Hosoya,Hayward}.  It is therefore of
some importance to understand how good these approximations are.

If the boundary values of real tunneling solutions are the maxima
of the Hartle-Hawking wave function, then classical and semiclassical
approximations may be reasonably reliable.  In particular, the classical
signature-changing solutions of the Einstein equations are precisely the
real tunneling geometries, and the interesting predictions based on these
solutions \cite{Hayward} presuppose that they dominate the quantum wave
function.  Similarly, investigations of the sum over topologies
\cite{entropy,sum,Hosoya} have been based on the properties of real
tunneling solutions, and may be invalid if the peaks of the wave function
lie elsewhere.  On the other hand, if the real tunneling contributions
are {\em not\/} the maxima of the Hartle-Hawking wave function, this would
have interesting implications as well:  the wave function would then have
no maxima, and would presumably be dominated by behavior near some boundary
of the space of spatial geometries.

This paper has not answered the question of whether the Hartle-Hawking
wave function has maxima.  We have seen, however, that if any such maxima
exist, they must come from the contributions of real tunneling geometries.
In the simple model of quantum gravity in three spacetime dimensions, we
have found some reasonably strong evidence that these extrema are indeed
maxima, but in realistic four-dimensional gravity the question remains
open.
\vspace{2ex}
\newpage
\begin{flushleft}
\large\bf Acknowledgements
\end{flushleft}

I would like to thank Joel Hass and Geoff Mess for helpful discussions of
three-dimensional geometry and topology.  This research was supported in
part by the U.S.\ Department of Energy under grant DE-FG03-91ER40674.

\appendix
\section{\ }

In this appendix, we fill in some of the details in the derivation of
equation \rref{3x8} from \rref{3x7}.  Note first that the tangent
directions to $\partial M$ are defined without reference to the metric
on $M$, so the direction of normal $n_a$ is also independent of the
metric.  We can therefore write
\beq
\delta n_a =  \alpha n_a = {1\over2}(\gamma_{bc}n^bn^c)n_a
   \quad \hbox{on $\partial M$} .
\label{a1}
\eeq

To proceed further, it is useful to partially fix the gauge.  An
infinitesimal diffeomorphism corresponds to a transformation
\beq
\gamma_{ab}\rightarrow\gamma_{ab} + \nabla_{\!a}\xi_b + \nabla_{\!b}\xi_a
   \ , \qquad n^a\xi_a = 0 \quad \hbox{on $\partial M$} ,
\label{a2}
\eeq
and metrics that differ by such transformations are physically equivalent.
We can use this freedom to set
\beq
n^a\gamma_{ab}h^b{}_c = 0 \quad \hbox{on $\partial M$} ,
\label{a3}
\eeq
implying that
\beq
\gamma_{ab}n^b = 2\alpha n_a \quad \hbox{on $\partial M$} .
\label{a4}
\eeq
The remaining diffeomorphisms are now restricted to those that preserve
\rref{a3}, i.e.,
\beq
n^a\xi_a=0 \ ,\quad \nabla_{\!n}(\xi_a h^a{}_c) = 0 \quad\hbox{on
   $\partial M$}.
\label{a5}
\eeq
(We have used the fact that $K_{ab} = 0$, which together with \rref{3x2}
implies that $\nabla_{\!a}n_b=0$ on $\partial M$.)

We could fix the remaining freedom by using the fact that
\beq
\beta_a = \nabla^b(\gamma_{ab} - {\textstyle{1\over2}}g_{ab}\gamma)
   \rightarrow \beta_a + (\Delta_g\xi_a + R_a{}^b\xi_b) .
\label{a6}
\eeq
Given the boundary conditions \rref{a3}, ${\cal D}_a{}^b =
\Delta_g\delta_a{}^b + R_a{}^b$ is a self-adjoint elliptic operator.
With appropriate assumptions of smoothness, this means that the equation
${\cal D}_a{}^b\xi_b = -\beta_a$ can be solved, thus transforming $\beta_a$
to zero, unless $\beta_a$ is itself a zero-mode of ${\cal D}_a{}^b$.  Such
zero-modes are rare: in the case of empty space with a negative cosmological
constant, for instance, the eigenvalues of $\cal D$ are strictly negative.
To postpone dealing with the exceptional cases when $\cal D$ has zero-modes,
however, we shall not yet impose any condition on $\beta_a$.

We can now evaluate the variation of the extrinsic curvature $K_{ab}$:
\begin{eqnarray}
\delta K_{ab} &=& h_a{}^c\nabla_{\!c}\delta n_b
   - {\textstyle{1\over2}}h_a{}^cn^d(\nabla_{\!c}\gamma_{bd} +
   \nabla_{\!b}\gamma_{cd} - \nabla_{\!d}\gamma_{bc}) \nonumber\\
   &=& -h_a{}^c\nabla_{\!b}(\alpha n_c) +
  {\textstyle{1\over2}}h_a{}^c\nabla_{\!n}\gamma_{bc}
   = {\textstyle{1\over2}}h_a{}^c\nabla_{\!n}\gamma_{bc} ,
\label{a7}
\end{eqnarray}
where the fact that $\nabla_{\!a}n_b=0$ on $\partial M$ has been used
repeatedly.  Thus
\beq
\gamma^{ab}\delta K_{ab} =  {\textstyle{1\over2}}
  (\gamma^{ab} - n^an_c\gamma^{cb})\nabla_{\!n}\gamma_{ab}
  = {\textstyle{1\over2}}\gamma^{ab}\nabla_{\!n}\gamma_{ab} -
  2\alpha\nabla_{\!n}\alpha ,
\label{a8}
\eeq
and
\beq
\delta K = g^{ab}\delta K_{ab} = {\textstyle{1\over2}}\nabla_{\!n}\gamma
  - \nabla_{\!n}\alpha .
\label{a9}
\eeq
Note also that
\beq
h_{ab}\gamma^{ab} = \gamma - n_an_b\gamma^{ab} = \gamma - 2\alpha .
\label{a10}
\eeq

Inserting these results into equation \rref{3x7}, we find that
\beq
32\pi G \delta^2 I_E = \int_{\partial M}d^{n-1}\!x\,\sqrt{h}\!\left[
  \gamma^{ab}\nabla_{\!n}\gamma_{ab} - \gamma\nabla_{\!n}\gamma
  + 2\gamma\nabla_{\!n}\alpha + 2\alpha\nabla_{\!n}\gamma
  - 8\alpha\nabla_{\!n}\alpha
  \right] .
\label{a11}
\eeq
This equation can be further simplified by noting that
\begin{eqnarray}
n^a\beta_a = n_a\nabla_{\!b}\gamma^{ab} &-& {1\over2}\nabla_{\!n}\gamma
 = n_an_b\nabla_{\!n}\gamma^{ab} + n_ah_b{}^c\nabla_{\!c}\gamma^{ab}
  - {1\over2}\nabla_{\!n}\gamma \nonumber\\
 &=& 2\nabla_{\!n}\alpha  + h_b{}^c\nabla_{\!c}\,(2\alpha n^b)
  - {1\over2}\nabla_{\!n}\gamma
 = 2\nabla_{\!n}\alpha - {1\over2}\nabla_{\!n}\gamma ,
\label{a12}
\end{eqnarray}
leading directly to equation \rref{3x8}.

\end{document}